# Tailoring Architecture Centric Design Method with Rapid Prototyping

Nitish M. Devadiga

*Carnegie Mellon University, Pittsburgh, PA, USA

*Datarista Inc., Providence, RI, USA

ndevadig@alumni.cmu.edu, ndevadiga@datarista.com

*Abstract*

**Many engineering processes exist in the industry, text books and international standards. However, in practice rarely any of the processes are followed consistently and literally. It is observed across industries the processes are altered based on the requirements of the projects. Two features commonly lacking from many engineering processes are, 1) the formal capacity to rapidly develop prototypes in the rudimentary stage of the project, 2) transitioning of requirements into architectural designs, when and how to evaluate designs and how to use the throw away prototypes throughout the system lifecycle. Prototypes are useful for eliciting requirements, generating customer feedback and identifying, examining or mitigating risks in a project where the product concept is at a cutting edge or not fully perceived. Apart from the work that the product is intended to do, systemic properties like availability, performance and modifiability matter as much as functionality. Architects must even these concerns with the method they select to promote these systemic properties and at the same time equip the stakeholders with the desired functionality. Architectural design and prototyping is one of the key ways to build the right product embedded with the desired systemic properties. Once the product is built it can be almost impossible to retrofit the system with the desired attributes. This paper customizes the architecture centric development method with rapid prototyping to achieve the above-mentioned goals and reducing the number of iterations across the stages of ACDM.**

## I. INTRODUCTION

In software engineering industries various standards are adopted. These standards most often emphasize on specific development paths and suggest a specific style of engineering independent of the project or product being developed. Some examples of such processes are the traditional Waterfall model [11], Boehm's Spiral Model for software development [10] and MIL-STD-498 [12].

In organizations, customizing or selecting a suitable engineering process is a common activity, while Defense projects tend to follow and prescribe standards in a somewhat stricter fashion. Even then Military standards undergo revision every few years because of optimization, rework and need for additional flexibility [1].

The primary objective of this paper is to describe the advantages of investing more time in the requirements and analysis phase of product development for a better high quality deliverable and then to achieve this propose a tailored version of architecture centric development method with rapid prototyping. Several activities are identified as useful to the engineering process i.e. the process by which the entire system including the software, hardware, systemic properties and components are designed and developed [1]. These activities are identified by requirement elicitation and by tailoring the ACDM process.

Several points were significant in defining the process outlined in this paper:

1. The author while adopting Personal Software Process(PSP) in conjunction with other students in Carnegie Mellon University observed that if valuable time is invested in the requirements and design then amount of time spent in development is relatively less and the quality of the product is much higher with a major reduction in the number of defects.

2. On adopting ACDM over a period of three months in few small-scale projects and assignments it was observed that any changes in the requirements by the customer at the design stage resulted in re-iterating the process from stage 1. This was un-productive and a time-consuming task. This is mitigated by adopting rapid prototyping in the initial stage of ACDM.

3. As observed industry wide and in the authors experience, customers are not fully aware of what is required from the product. Rapid prototyping seemed to be the right choice to mitigate this.

4. While performing proof of concept or experiments rapid prototyping was one of the key methods as after each cycle we could identify progress, make note errors and mistakes and resolve in the next cycle. However, one important thing to note is prototypes should not be converted to final products as they lack the quality and systemic properties desired by the production system [8].

5. In a traditional engineering processes, total estimates

made during the in the initial stage are highly inaccurate. ACDM with rapid prototyping mitigates this fact, as through solidifying the requirements in the initial stage better estimates can be made based on the requirements, resource availability, skill and historical data. Also, ACDM identifies the initial stage

## II. VERIFICATION

As discussed earlier it is observed industry wide that if valuable time is spent in requirement analysis and design phase it improves the overall agility and quality of project.

To validate this theory author along with other students of Masters of Software Engineering program at Carnegie Mellon University had adopted Personal Software Process (PSP) [4] on a set of assignments. Assignments were to be completed by gradually applying the PSP process with principle emphasis on understanding the problem statement thoroughly (Requirement analysis), estimation and designing the solution for the problem statement. Some of the key observations from this experiment are as described:

1. Time spent on planning and design has increased gradually from the 1st assignment to the last, to achieve high quality deliverable. The linear trend line identifies the increase in planning time.

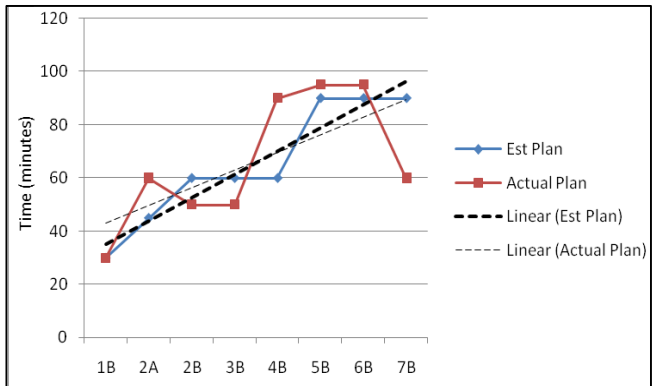

*Figure 1  Increase in planning time from 1B to 7B*

2. From the below graph we can measure the gradual increase in design time.

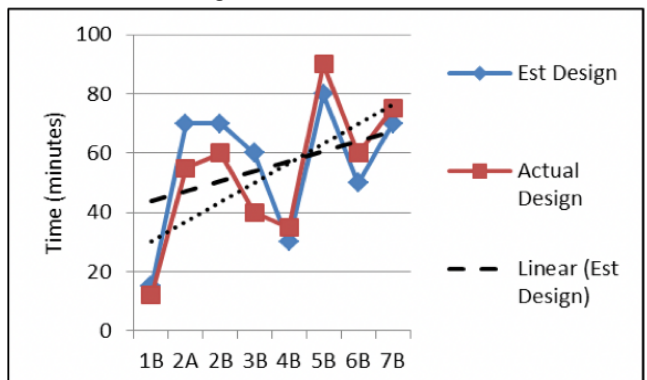

*Figure 2 Increase in design time from 1B to 7B*

3. This increased amount of time allocated for planning and design has reduced the time required for coding i.e. development time, as thorough analysis is done on project startup; this approach has also decreased the number of defects. Here we also must assume that the students are proficient in coding. Thus, this factor though considered for the reduction in coding time does not play as major role.

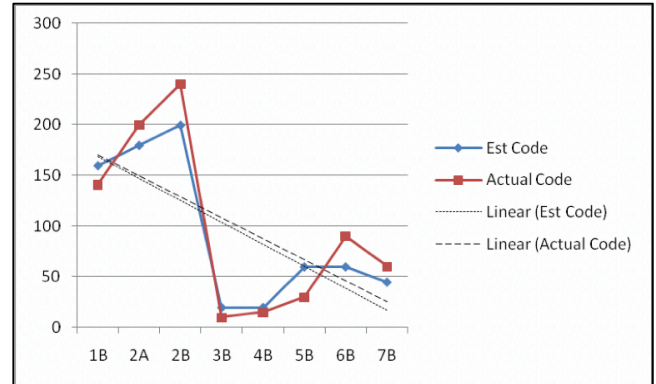

*Figure 3 Decrease in development time from 1B to 7B*

4. Based on the values in Table 3 in appendix on plotting defect/program graph we can view that the number of defects is in decreasing order.

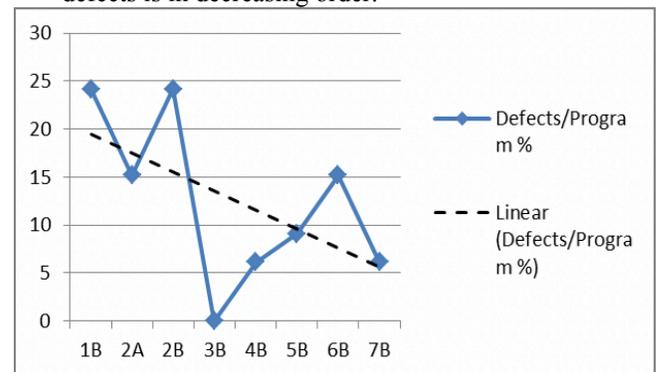

*Figure 4 Decrease in Defects from 1B to 7B*

From the above analysis, we can be sure that if quality time is invested in performing thorough requirement analysis and design then it improves the overall quality of the project by reducing the development time and the number of defects.

## III. REVISED ENGINEERING PROCESS

### A. RAPID PROTOTYPING

A prototype is an executable model of a system that accurately reflects a chosen subset of its properties, such as display formats, computed results, or response times. Prototypes are useful for formulating and validating requirements, resolving technical design issues, and supporting computer-aided design of both software and hardware components of proposed systems [5].

There are two types of prototypes: 1. Throw away or rapid prototype: This is mainly used for requirement elicitation, to equip the users with the feel of the system. The key attribute of this is the speed in which the model is provided. Once their purpose is served the prototype is discarded or just used for reference and not as a final product. [6]

2. Evolutionary prototype: To build the system through

continuous iteration and keep adding features as per requirement till a stable final product is achieved.

Rapid prototyping refers to the capability of creating a prototype with significantly less effort than it takes to produce an implementation for operational use. In which emphasis is placed on developing prototypes early in the software life cycle to understand the customer's requirement, permit early feedback and analysis in support of the software process [5].

Rapid prototyping is packaged in such a way that it can be embedded into several existing design processes. In this paper, rapid prototyping is embedded in stage 1 of ACDM and is informally used to carry out experiments in stage 6 of ACDM.

### B. ARCHITECTURE CENTRIC DESIGN METHOD (ACDM)

The ACDM is a design method for organizations and teams building software intensive systems. A system that in all respects depends upon software to provide its specified services is a software intensive system [2].

It uses the architectural design to systematically explore and refine the design, refine the architectural drivers, and mitigate technical risks. Technological issues are noted early and addressed well before they can impact cost and schedule during detailed design and implementation [2].

Non-functional requirements such as performance, modifiability, and availability etc. are as important as functional requirements. These systemic properties strongly influence the design of the system. Architects need to constantly even these concerns with the structures they select through design to promote the properties and provide the functionality required by the stakeholder community. If these systemic properties are not designed into the architecture it can be impossible to retrofit systems with these properties once they are built [3]. These concerns are taken care by ACDM. Below table describes the various stages of ACDM.

| | | |
|---|---|---|
| Period of Uncertainty | Stage 1: | Establish the Architectural Drivers |
| | Stage 2: | Establish Project Scope |
| | Stage 3: | Create/Refine the Architecture |
| | Stage 4: | Evaluate the Architecture |
| | Stage 5: | Production Go/No-Go Decision |
| | Stage 6: | Plan and Execute Experiments |
| Period Of Certainty | Stage 7: | Production Planning |
| | Stage 8: | Production |

*Figure 5 - 8 Stages of ACDM*

To ensure that a system possesses the necessary systemic properties, functionalities and adheres to the constraints as desired by client, it is important to elicit all the requirements (done using rapid prototyping) and ensure system architectures are designed early and used as a guideline for design and implementation (done using ACDM).

### C. ACDM with RAPID PROTOTYPING

The fundamental part of this process is to allow the pre-engineering phase to form a formal part of the engineering process to:

1. Allow the prototypes to assist the customer in requirements identification

2. Gain important technical insight into the product. Informally identify the architectural drivers.

3. Any additions/updates in the requirements by the customer at the design stage resulted in re-iterating the process from stage 1. This was un-productive and a time consuming and labor intensive task. This is mitigated by adopting rapid prototyping in the initial stage of ACDM as it would solidify the requirement in the initial stage.

4. Ability to provide approximate estimates after stage 1 by completely understanding the functional and non- functional requirements.

5. Tighten estimation of development time and cost.

6. Research novel concepts to help customer and developer analyze the system interactions, record

close system calls, identify system risks or issues.

Figure 6 illustrates the revised engineering process as referenced in the Introduction.

| | | |
|---|---|---|
| Period of Uncertainty | Stage 1: | Rapid Prototyping - Requirement Elicitation and Informally identifying Architectural drivers |
| | Stage 2: | Establish Project Scope |
| | Stage 3: | Create/Refine the Architecture |
| | Stage 4: | Evaluate the Architecture |
| | Stage 5: | Production Go/No-Go Decision |
| | Stage 6: | Rapid Prototyping - Carry out Experiments |
| Period Of Certainty | Stage 7: | Production Planning |
| | Stage 8: | Production |

*Figure 6 Tailored ACDM with Rapid Prototyping*

### Period of Uncertainty

### Stage 1: Rapid Prototyping requirements elicitation and informal architectural driver's identification

Analysts and development team interact with system stakeholders to discover functional and non-functional requirements, define and document the interfaces and architectural drivers. Using this information, a high level prototype is built immediately and demonstrated to the customer. Based on the customers feedback the prototype is iteratively modified till the following attributes are obtained as desired by the customer.

1. Functional requirement: Describing what the system must do.

2. Technical constraints: Technical constraints are technical decisions that are mandatory for the team to incorporate such as the use of legacy software, operating system etc.

3. Business constraints: Business constraints are related to the functional requirements of the product; these do not specify a technical approach but strongly imply specific technical properties.

4. Non-functional requirements: Systemic properties such as modifiability, availability, performance that the product must possess.

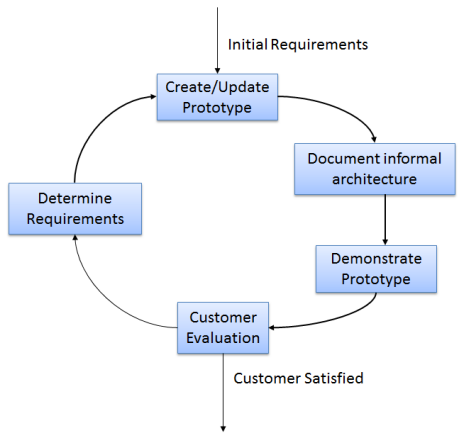

*Figure 7 Rapid Prototyping*

Here feedback from customers in the previous stage is collated and analyzed to produce a set of formal user requirement specification.

It is difficult to extract all the requirements logically and consistently all in one operation. Reasonably accurate prototypes can be created via rapid prototyping method that produces a series of related prototypes to converge on a consensus about the requirements [6].

Rapid prototyping may produce a series of prototypes, $P[i]$ = 1, 2, ..., n, ... All these prototypes are increasingly accurate approximations of the envisioned system. Information gathered from user feedback, analyzing and criticizing $P[i]$ is used to build $P[i + 1]$. Here the user feedback, analysis, architectural changes and other modification done to prototype $P[i]$ is represented as $\Delta P[i]$, thus we can represent this as, $P[i] + \Delta P[i] \approx P[i+1]$ Where $P[i+1]$ is a better approximation i.e. better prototype of system than $P[i]$.

This process should continue till the value of $\Delta P[i]$ is small enough that is acceptable to the available budget, resources and satisfies the client requirements. If we can assume convergence of such a series based on human cognitive ability, then the final prototype P would be,

P = limit $S[i]$ when i > ∞ Practically since the resources are limited, an integer n = N must be chosen such that $P[N]$ approximates prototype P and the differences are $\Delta P[N]$ minute enough to be acceptable relative to the schedule client requirements, available resources and cost [6]. Here the value of N is very crucial as it determines the number of times the system has been modified. The task of constructing the sequence $P[i]$ = 1, ...., N could be an exhaustive process and requires continuous interactions with customers, however once the customer has agreed to the prototype and the architectural attributes we can be sure that the requirements are clear and the right product is going to be built. The key benefits of performing this activity is that by the end of this cycle from $P[i]$ = 1, ..., N we will have a tangible throw away prototype agreed by the customer with clear requirement specification. As the architectural attributes are informally documented at every stage and used for requirement elicitation with customer; we will have a well-documented informal architectural specification. Architectural specification should not be postponed after the requirement phase is completed as it is one of the key drivers to requirement elicitation and analysis. It helps in managing customer expectations, identifying constraints, identifying business drivers, and narrowing down the design solution space which helps in specifying project scope and plan. Stage 2: Establish project scope

1. The team organizes, consolidates and refines architectural drivers.

2. Team creates initial plan for the period of uncertainty. This plan lists the expected number of iterations in stages 3, 4, 5 and 6.

As we have performed rapid prototyping in stage 1, with the help of the prototype we can easily organize and consolidate the architectural drivers.

At this point with the clear requirement specification and informally identified architectural drivers we can approximately estimate the project.

Also, we should note that the number of iterations for stages 3, 4, 5 and 6 would be reduced as the requirement is clear and from the stage 1 what is supposed to be done and what is required for this project.

### *Stage 2: Establish project scope*

1. The team organizes, consolidates and refines architectural drivers.

2. Team creates initial plan for the period of uncertainty. This plan lists the expected number of iterations in stages 3, 4, 5 and 6.

As we have performed rapid prototyping in stage 1, with the help of the prototype we can easily organize and consolidate the architectural drivers.

At this point with the clear requirement specification and informally identified architectural drivers we can approximately estimate the project.

Also, we should note that the number of iterations for stages 3, 4, 5 and 6 would be reduced as the requirement is clear and from the stage 1 what is supposed to be done and what is required for this project.

### *Stage 3: Create/Refine the Architecture*

At stage three the development team designs the architecture. ACDM prescribes continuous evaluation and refinement of the architecture until deemed to be fit. The biggest problem thus far faced by the ACDM process is getting architects to design the notational architecture quickly without lengthy deliberation [2]. Architects will spend more time designing the perfect architecture, the key concept of ACDM is not to spend much time trying to design perfect architecture as it would be evaluated in stage 4 and the output of the evaluation will guide its refinement.

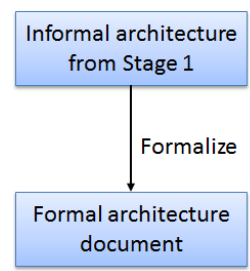

However, it is observed as a common practice that the amount of time spent in the creating the architecture document for the 1st time is very high. This initial cost is reduced by using the informal architecture document created in stage 1 during the rapid prototyping process. Formalizing an informal document is much less time consuming rather than building it from scratch. Also since the document is built using informal architecture document which is a product of number of iterations in stage 1, the architecture document created at this stage would be stable and less prone to changes. This would in turn reduce the number of iterations in stage 3, 4, 5, and 6.

### Stage 4: Evaluate the Architecture

Evaluate the architecture and determine the state and fitness of the architectural design. Initial evaluations can be conducted with internal stakeholders and later evaluations should include both internal and external stakeholders.

### Stage 5: Production Go/No-Go Decision

Team analyzes the information from stage 4 and decided whether the architecture is ready to go into production which is Go. If the architecture is not ready and needs more refinement, then a No-Go decision is made and architecture and to be refined further. At this stage, it can be possible that some parts are production ready and some are not so this decision can be carefully made considering the dependencies and some parts can go into production while others are reiterated for further refinement.

### Stage 6: Plan and Execute Experiments

With the issues identified in stage 4 experiments are planned. Issues provide information that can be used to refine the architecture. At this stage, rapid prototyping is used to iterate the architecture design, not the product. ACDM describes a rigorous experimentation process and provides templates for planning experiments and guiding teams in experimentation. Here one key point to remember is that the prototype created should be a throw away prototype. As here throw away rapid prototyping experiments is used in testing concepts, technologies, reduce learning curves, refining architectural drivers, and so forth.

### Period of Certainty

### Stage 7: Production Period of Planning

Once the architectural design is refined thorough the iteration in the above stages from 3 to 6 i.e. period of uncertainty, the team moves into production stages i.e. period of certainty. At this point as the architectural design is established high fidelity estimates and production schedules can be made.

### Stage 8: Production

The team performs the detailed design and implements the product.

1. Design: Development team can use the architectural design document to create a detailed design document. Team can choose any method to create detailed design.

2. Construction and unit test: Elements of the system will be built and tested.

3. Integration and system test: Architectural elements are integrated and tested.

### D. OBSERVATIONS

ACDM requires iteration around the entire design process in case an organization wants to make changes or add to the requirements. The system architecture then must be defined again in stages 3, 4, 5, and 6. This drawback is mitigated by this above proposed methodology of performing ACDM with rapid prototyping. Since all the requirements and architectural drivers are comprehensively captured in stage 1 using rapid prototyping. This reduces the amount of overhead of iterating through the whole process again and as we know that the customer requirements keep changing intermittently this approach seems to be the right choice.

Based on the authors previous experience and the data collected while experimenting with this process it was observed that prototype 1 required the maximum time, in the next two iterations the time spent almost reduced by 60%.

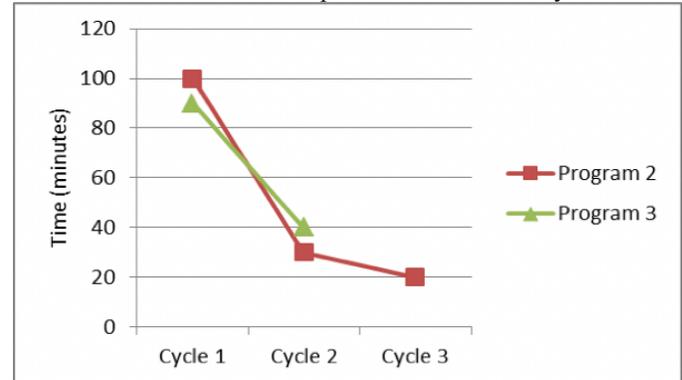

*Figure 8 Time taken for Rapid Prototyping*

From the figure 6 we can see the rise in the amount of time spent on requirement and design.

In this experimentation result of program 2 and program 3 is compared to that of program 1. Program 1 was not developed using the above methodology and program 2 and 3 use the above defined process. Also, we can safely assume that all the programs were of almost similar difficulty and the developers are proficient in coding.

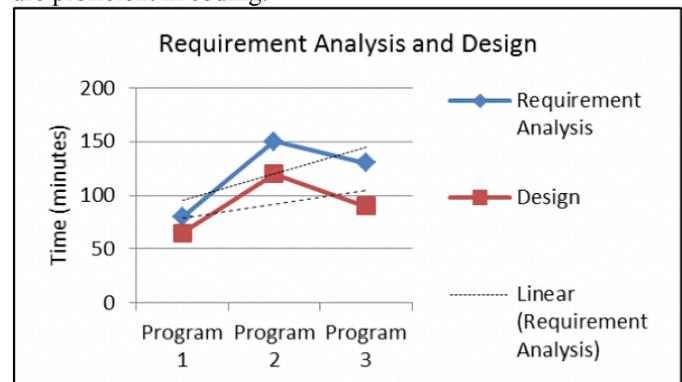

*Figure 9 Time taken Requirements and Design*

The informal architectural document helped in formalizing and refining the architectural document. The time taken for development for project two and three has reduced compared to project one.

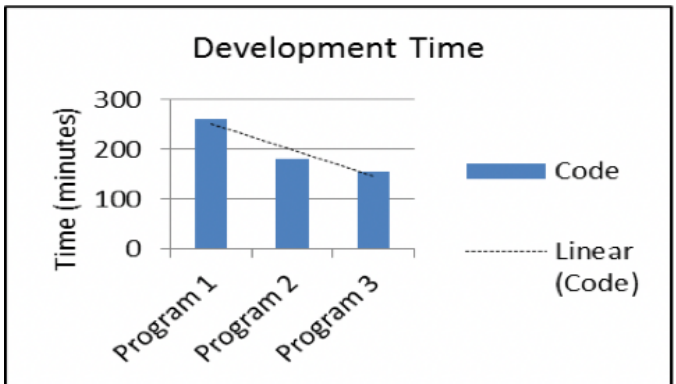

*Figure 10 Time taken Development*

The number of defects in project two and three has gradually decreased.

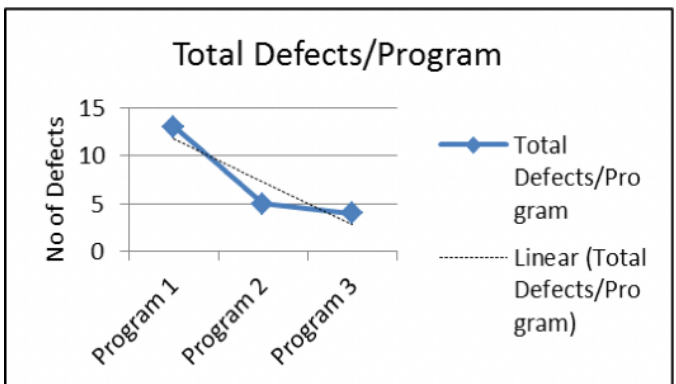

*Figure 11 Defects per Program*

From Figure 12 we can analyze the number of defects per phase. We can observe that the number of defects in program 2 and 3 to which the above process was applied has lesser number of defects compared to program 1.

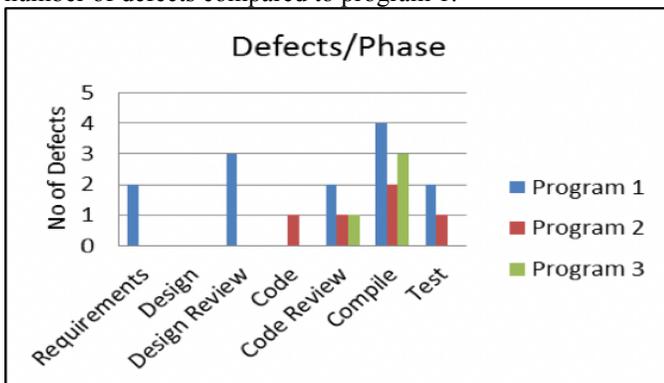

*Figure 12 Defects per Phase*

## IV. ANALYSIS

### A. SCALABILITY

The tailored process presented in the paper is applicable to more than software development. It can be envisioned to be followed for many software/hardware projects.

One of the key benefits of adopting such a process is the requirement to produce secure and robust products to a strict deadline. This may seem like an essential target for any project, however typically, the literature [13, 14] and practical experience indicates deadline slippage is common, especially in cases where risks are not mitigated. Often the slippage occurs late in the development phase or more likely during the debugging phase.

In this process once we have completed the requirements and design both the customer and the user have a clear understanding of the system and the technical issues in designing the end system. The process is viable based on the:

- Scale aka life span of the project
- Available resources
- Available time

A smaller project can follow a similar path however it would be expected to limit any introductory research or development to keep the project small scale and avoid increase in cost

Mainly large projects would benefit from this type of process. Though they may be complex or have more amounts of risks but early and ongoing prototyping heavily assists in mitigating technical as well as requirements related to risks within project.

### B. PROTOTYPING

Prototyping is said to be one of the useful tools for requirement elicitation. While prototyping is useful to obtain large amount of reuse from prototyped code, it is more useful to obtain a rapidly prototyped system, so that the functionality and user interface of the system can be demoed and accessed by the customer.

### C. MANAGEMENT ISSUES

The process mentioned in this paper can be applied to arbitrary team sizes. However, the routine management issues must be addressed, such as which teams are working on which part of the system. In large team's communication is very important. For example, early communication risks mitigation results.

Communication must be established by conducting meetings, reviews, status meetings in a timely manner to ensure all teams are in sync.

### D. LESSONS LEARNT

The key advantage of using this process is when the size, complexity and technical risk increases in a project. Any amount of technical insight gained prior to design stages is of very high value. The cardinal purpose of carrying out such experiments is to test the value and implementation feasibility of new ideas.

In the authors experience it is common that rapid prototyping or any other prototyping method, experimentation results in systems that are unusable. However, the value of the prototyping and experiments carried out comes from the analysis of the failure as an aid in identifying the nature of poor results and highlighting areas that require attention and effort.

The results of failed or successful experiments equips in accurately estimating the design and development profile of the system. This may prove valuable in those rare instances when schedule, resource, and cost are all critical!

## V.  CONCLUSION

In conclusion, the paper clearly states the utility of rapid prototyping in the architecture centric design method. By thoroughly applying rapid prototyping techniques the time impact on a project can be reduced as it can reduce the amount of time iterating the whole or sub process [9].

We can now argue that eliciting better specifications in this way is a means of reducing maintenance costs. Better specifications imply more content users and fewer change requests. This does not mean that maintenance cost will disappear, only that the burden is mitigated. It is widely reported that up to 50% of maintenance requests, principally those which occur in the nine months or so immediately following delivery result from misspecification [7]. It cannot be more emphasized that this sort of maintenance effort is unproductive labor. Reducing it by any amount will benefit the organization by giving it more time to build features and focus on newer customer requirements.

By customizing the:

- Extent of prototyping
- Formality of design and review phase
- Early involvement with the customer in eliciting requirements,

We can think of a more precise profile of the development phases, at the same time enabling project managers estimate a more accurate schedule for project development.